\documentstyle[preprint,prc,aps,epsf]{revtex}
\begin{document}
\draft
\title{Coulomb Distortion Effects for Electron or Positron Induced $(e,e')$ Reactions in the Quasielastic
Region}
\author{K.S. Kim}
\address{Department of Physics and Institute of Basic Science,
Sung Kyun Kwan University , Suwan, 440-746, Korea}
\author{L.E. Wright and D.A. Resler}
\address{
Institute of Nuclear and Particle Physics,
Department of Physics and Astronomy, Ohio University, Athens,
Ohio 45701}

\maketitle
\begin{abstract}
In response to recent experimental studies we investigate Coulomb
distortion effects on $(e,e')$ reactions from medium and heavy
nuclei for the case of electrons and positrons. We extend our
previously reported full DWBA treatment of Coulomb distortions to
the case of positrons for the $^{208}Pb(e,e')$ reaction in the
quasielastic region for a particular nuclear model.  In addition,
we use previously reported successful approaches to treating
Coulomb corrections in an approximate way to calculate the Coulomb
distortion effects for $(e,e')$ reactions for both electrons and
positrons for the case of a simple nuclear model for quasielastic
knock-out of nucleons. With these results in hand we develop a
simple {\em ad-hoc} approximation for use in analyzing
experiments, and discuss methods of extracting the ``longitudinal
structure function" which enters into evaluation of the Coulomb
sum rule. These techniques are generally valid for lepton induced
reactions on nuclei with momentum transfers greater than
approximately 300 $MeV/c$.
\end{abstract}
\narrowtext

\section{Introduction}

 A persistant problem in using electron scattering for
investigating nuclear structure and nuclear properties,
especially in the quasielastic region, is the large static
Coulomb field of medium and heavy nuclei.  The presence of the
static Coulomb potential (of order $25$ MeV at the surface of the
$^{208}Pb$ nucleus) invalidates one of the primary attributes of
electron scattering as usually presented.  Namely that in the
electron plane-wave Born approximation, the cross section can be
written as a sum of terms each with a characteristic dependence
on electron kinematics and containing various bi-linear products
of the Fourier transform of charge and current matrix elements.
That is, various structure functions for the process can be
extracted from the measured data by so-called Rosenbluth
separation methods. The trouble with this picture is that when
Coulomb distortion of the electron (or positron) wavefunctions
arising from the static Coulomb field of the target nucleus is
included exactly by partial wave methods, the structure functions
can no longer be extracted from the cross section, even in
principle.

In the early 90's, Coulomb distortion for the reactions $(e,e')$
and $(e,e'p)$ in quasielastic kinematics was treated exactly by
the Ohio University group\cite{jin1,yates,jin2,zhang,jin3}  using
partial wave expansions of the electron wavefunctions. Such
partial wave treatments are referred to as the distorted wave
Born approximation (DWBA) since the static Coulomb distortion is
included exactly by numerically solving the radial Dirac equation
containing the Coulomb potential for a finite nuclear charge
distribution to obtain the distorted electron wave functions.
While this calculation permits the comparison of  nuclear models
to measured cross sections and provides an invaluable check on
various approximate techniques of including Coulomb distortion
effects, it is numerically challenging and computation time
increases rapidly with higher incident electron energy.
Furthermore, the initial computer codes did not include the
option of calculating positron induced reactions in an obvious
manner although only the sign of the Coulomb distortion term in
the Dirac equation needed to be changed.  And, as noted above, it
was not possible to separate the cross section into various terms
containing the structure functions and develop insights into the
role of various terms in the transition charge and current
distributions.

In our DWBA investigations of $(e,e')$ and $(e,e'p)$ reactions in
the quasielastic region, we  used a relativistic treatment based
on the $\sigma-\omega$ model for the nucleons involved. In
particular, for the $(e,e'p)$ reaction we use a relativistic
Hartree single particle model for a bound state\cite{horo} and a
relativistic optical model for an outgoing proton\cite{hama}
combined with the free space relativistic current operator
$J^{\mu} = \gamma^{\mu} + i {\frac {\kappa} {2 M}}
\sigma^{\mu\nu}\partial_\nu$. For the $(e,e')$ case we solve for
the continuum nucleon wavefunctions using the real bound state
potential so as to maintain current conservation.  Using these
models, we compared our DWBA calculations with experimental data
measured at various laboratories for $(e,e')$\cite {jin1,yates},
and for $(e,e'p)$ \cite{jin2,zhang,jin3} and have found excellent
agreement with the data. We concluded that the relativistic
nuclear models are in excellent agreement with the measured data
and note that we do not need to invoke meson exchange effects and
other two-body terms in the current that are necessary in a
Schr\"{o}dinger description that uses a non-relativistic
reduction of the free current operator \cite{gent}.  However,
other investigators use other nuclear models and our elaborate
DWBA code can not be easily modified to include different
transition currents.

To avoid the numerical difficulties associated with DWBA analyses
at higher electron energies and to look for a way to still define
structure functions, our group \cite{kim1,kim2,kim3} developed an
approximate treatment of the Coulomb distortion based on the work
of Knoll\cite{knoll} and the work of Lenz and
Rosenfelder\cite{lenz}. We were able to greatly improve some
previous attempts along this line\cite{giusti,trani} where
various additional approximations were made which turned out not
to be valid.   The essence of the approximation is to calculate
the four potential $A_\mu$ arising from the lepton four current in
the presence of the static Coulomb field of the nucleus.  This is
possible for momentum transfers greater than approximately $300 $
MeV/c in a limited spatial region which we take to be of order $3
R$ where $R$ is the nuclear charge radius.  The Coulomb
distortion is included in the four potential $A_\mu$ by the
elastic scattering lepton phase shifts and by letting the
magnitude of the lepton momentum include the effect of the static
Coulomb potential.  This last step leads to an $r$-dependent
momentum. A key result of our approximation method is that the
separation of the cross section into a ``longitudinal'' term and
a ``transverse'' term is still possible.

We compared our approximate treatment of Coulomb distortion (which
we will designate as {\em approximate DW}) to the exact DWBA
results for the reaction $(e,e'p)$ and found good agreement (at
about the 1-2$\%$ level) near the peaks of cross sections even
for heavy nuclei such as $^{208}Pb$. With an improved
parametrization of the elastic scattering electron phase shifts
\cite{kim3}, we achieve quite good agreement away from the peaks
in the cross sections. Using this approximate DW treatment of
Coulomb distortions for the inclusive $(e,e')$ reaction in much
more difficult numerically since  the direction of the outgoing
nucleon has to be integrated over, and all the nucleons in the
nucleus have to be knocked out. Therefore, we sought even more
severe approximations in order to obtain a simple {\em ad-hoc}
method of calculating the structure functions for $(e,e')$
reactions.  In our earlier work, we found it necessary to use
different {\em ad-hoc} procedures for the longitudinal and
transverse terms\cite{kim2}, although our investigation of the
{\em ad-hoc} procedure for the longitudinal terms was hindered by
the fact that the longtitudinal contributions to the total cross
section are usually considerably less than $50 \%$ and thus we did
not have great sensitivity to the Coulomb corrections for the
longitudinal structure function.   In this paper we will use a
simple non-relativistic {\em toy} model to calculate the Coulomb
corrections to the longitudinal structure function with our {\em
approximate DW} methods that we applied to $(e,e'p)$ and then
investigate the {\em ad-hoc} treatment of the longitudinal
structure function which is a key ingredient in investigating the
Coulomb sum rule.  After developing an improved {\em ad-hoc}
procedure using our {\em toy} model we compare it the the full
DWBA calculation which we have now extended to include positron
induced reactions.

\section{Approximate Treatment of Coulomb Distortion}

 Our approximate method of including the static Coulomb
distortion in the electron wavefunctions is to write the wave
functions in a plane-wave-like form\cite{kim2};
\begin{equation}
{\Psi}^{\pm}({\bf r})={\frac {p'(r)} {p}}\;e^{{\pm}i{\delta}
({\bf L}^{2})}\;e^{i\Delta}\;e^{i{\bf p}'(r){\cdot}{\bf
r}}\;u_{p}\;, \label{oldwv}
\end{equation}
 where the phase factor
$\delta({\bf L}^{2})$ is a function of the square of the orbital
angular momentum, $u_{p}$ denotes the Dirac spinor, and the local
effective momentum ${\bf p}'({\bf r})$ is given in terms of the
Coulomb potential of the target nucleus by
\begin{equation}
{\bf p}'({\bf r})=\left( \; p-{\frac{1}{r}} \int^{r}_{0} V(r)dr
\right ){\bf {\hat p}} \;. \label{lema}
\end{equation}
The $ad-hoc$ term $\Delta=a[{\bf {\hat p}}'(r){\cdot}{\hat r}]
{\bf L}^{2}$ denotes a small higher order correction to the
electron wave number which we have written in terms of the
parameter $a=-{\alpha}Z(\frac{16 MeV/c}{p})^{2}$. The value of 16
MeV/c was determined by comparison with the exact radial wave
functions in a partial wave expansion.  We have examined the
positron case ($Z\mapsto-Z$) and find that this parametrization
works equally well when compared to the exact radial positron
wave functions.

We calculate the elastic scattering phases and fit them to a
function of the square of the Dirac quantum number $\kappa$ used
to label the phase shifts. We then replace the discrete values of
$\kappa^2$ with the total angular momentum operator ${\bf J}^2$
which we subsequently replace by the orbital angular momentum
operator ${\bf L}^2$ since the low $\kappa$ terms where the
difference between $j$ and $l$ is significant contribute very
little to the cross section. Finally we replace the angular
momentum operator squared by its classical value $({\bf r} \times
{\bf p})^2$.  The removal of any spin dependence apart from what
is in the Dirac spinor $u_p$ is crucial for writing the cross
section as the sum of a longitudinal and a transverse
contribution.

Initially \cite{kim1} we fitted the phases $\delta_\kappa$ to a
quadratic function of $\kappa^2$ which worked reasonably well for
lower electron energies, but with the prospect of new higher
energy electron accelerators, we needed a fit to the phases that
will work at higher energies. In addition, we wanted to avoid
calculating all of the elastic phase shifts, particularly the
very high $\kappa$ values. We decided to make use of the fact
that the higher $\kappa$ phase shifts approach the point Coulomb
phases which have a simple analytical form at high energy. The
low $\kappa$ phases, corresponding to orbitals which penetrate
the nucleus, are linear in $\kappa^2$ which was the basis or our
initial parametrization. The difficult phases to fit correspond
to $\kappa$ values of order $pR$ which, from a classical point of
view, correspond to scattering from the nuclear surface region
and are known to make large contributions to the cross section.
We were able to find a parametrization of the elastic scattering
phases shifts in terms of $\kappa^{2}$ which has the correct
large $\kappa^2$ behaviour and becomes linear in $\kappa^2$ at
low angular momentum, and since we have the correct large
$\kappa$ behaviour, we need only calculate the exact scattering
phase shifts for $\kappa$ values up to order $pr$. After some
investigation \cite{kim3}, we found that the following
parametrization of elastic scattering phase shift describes the
exact phase shifts very well:
\begin{equation}
{\delta}({\kappa})=[a_0 + a_2\frac{{\kappa}^{2}}{(pR)^{2}}]
e^{-{\frac{1.4{\kappa}^{2}}{(pR)^{2}}}}-{\frac{{\alpha}Z}{2}}(1-e^{-{\frac{{\kappa}^{2}}{(pR)^{2}}}})
{\times}{\ln}(1+{\kappa}^{2})  \label{newph}
\end{equation}
where $p$ is the electron momentum and we take the nuclear radius
to be given by $R=1.2A^{1/3}-0.86/A^{1/3}$.  We fit the two
constants $a_0$ and $a_2$ to two of the elastic scattering phase
shifts ($\kappa=1$ and $\kappa=Int(pR)+5$). To a very good
approximation, $a_0=\delta(1)$ and $a_2=4\delta(Int(pR)+5)+\alpha
Z ln(2pR)$, where $Int(pR)$ replaces $pR$ by the integer just
less than $pR$. Note that this parametrization only requires the
value of the exact scattering phase shift for $\kappa=1$ and
$\kappa=Int(pR)+5$.  For this paper we have confirmed that this
same parametrization works equally well for the positron phase
shifts.

Using the new phase shift parametrization and the local effective
momentum approximation, we construct plane-wave-like wave
functions for the incoming and outgoing electrons.  Since the
only spinor dependence is in the Dirac spinor all of the Dirac
algebra goes through as usual and we end up with a M{\o}ller-like
potential given by,
\begin{equation}
A^{appro. DW}_{\mu} ({\bf r})={\frac {4{\pi}e} {q^2-\omega^2}
}e^{i[{\delta}_{i}(({\bf r}{\times}{\bf p}'
_{i}(r))^{2})+{\delta}_{f}(({\bf r}{\times}{\bf p}'_{f}(r))^{2})]}
e^{i({\Delta}_{i}-{\Delta}_{f})}e^{i{\bf q}'(r){\cdot}{\bf r}}
{\bar u}_{f}{\gamma}_{\mu}u_{i} \label{apppot}
\end{equation}
where the phase shift parametrization is given in Eq. 3 with
$\kappa^2$ being replaced by $({\bf r}\times {\bf p})^2$, the
parameter $\Delta$ is given following Eq.~\ref{lema}, and the
$r$-dependent momentum transfer is given by ${\bf q}'(r)={\bf
p}'_{i}(r)-{\bf p}'_{f}(r)$.

With this approximate DW four potential $A_\mu$ it is
straightforward to calculate the $(e,e'p)$ cross sections and
modified structure functions.  We showed \cite{kim3} that using
this new phase shift(see \cite{kim1,kim2} for details) we can
reproduce the full DWBA cross sections for $(e,e'p)$ from medium
and heavy nuclei very well.

\section{Application to the Inclusive Process}

In the plane wave Born approximation (PWBA), where electrons or
positrons are described as Dirac plane waves, the cross section
for inclusive quasielastic $(e,e')$ processes can be written
simply as
\begin{equation}
\frac{d^2\sigma}{d\Omega_e d\omega}= \sigma_{M} \{
\frac{q^4_\mu}{q^4}  S_L(q,w) + [ \tan^2 \frac{\theta_e}{2} -
\frac{q^2_\mu}{2q^2} ]  S_T(q,w) \} \label{pwsep}
\end{equation}
where $q_\mu ^2 = \omega^2-{\bf q}^2$ is the four-momentum
transfer, $\sigma _{M}$ is the Mott cross section given by
$\sigma_{M} = (\frac{\alpha }{2E} )^2  \frac{\cos^2
\frac{\theta}{2}}{\sin^4 \frac{\theta}{2}}$, and $S_L$ and $S_T$
are the longitudinal and transverse structure functions which
depend only on the momentum transfer $q$ and the energy transfer
$\omega$.   As is well known, by keeping the momentum and energy
transfers fixed while varying the electron energy $E$ and
scattering angle $\theta_e$, it is possible to extract the two
structure functions with two measurements.   As we will summarize
below, our approximate treatment of Coulomb distortions still
permit {\em Rosenbluth-like} separations but with Coulomb
corrections which require the use of models.

For the inclusive cross section $(e,e')$, the longitudinal and
transverse structure functions in Eq.~(\ref{pwsep}) are bi-linear
products of the Fourier transform of the components of the nuclear
transition current density integrated over outgoing nucleon
angles.  Explicitly, the structure functions for knocking out
nucleons from a shell with angular momentum $j_{b}$ are given by
\begin{eqnarray}
S_{L}(q,{\omega})&=&\sum_{{\mu}_{b}s_{p}}{\frac {{\rho}_{p}}
{2(2j_{b}+1)}} \int {\mid}N_{0}{\mid}^{2}d{\Omega}_{p} \\
S_{T}(q,{\omega})&=&\sum_{{\mu}_{b}s_{p}}{\frac {{\rho}_{p}}
{2(2j_{b}+1)}} \int
({\mid}N_{x}{\mid}^{2}+{\mid}N_{y}{\mid}^{2})d{\Omega}_{p}
\end{eqnarray}
where the nucleon density of states ${\rho}_{p}={\frac {pE_{p}}
{(2\pi)^{2}}}$, the $z$-axis is taken to be along ${\bf q}$, and
${\mu}_{b}$ and $s_{p}$ are the z-components of the angular
momentum of the bound and continuum state particles. The Fourier
transfer of the nuclear current $J^{\mu}({\bf r})$ is simply,
\begin{equation}
N^{\mu}=\int J^{\mu}({\bf r})e^{{\imath}{\bf q}{\cdot}{\bf
r}}d^{3}r.
\end{equation}
ans the continuity equation has been used to eliminate the
$z$-component ($N_{z}$) via the equation $N_{z}=-{\frac {\omega}
{q}}N_{0}$.

When we use our approximate M{\o}ller potential given in Eq.
(\ref{apppot}) , we also can separate the cross section into
longitudinal and tranverse components since as noted previously,
it the Dirac spinor structure that leads to this result.  However,
when we use the approximate electron four potential along with
current conservation to eliminate the z-component of the current
we run into a problem since the momentum transfer ${\bf
q}^{\prime}$ depends on r both in magnitude and direction. In
addition, the phase factors depend on ${\bf r}$. To avoid
generating additional terms we assume the direction of ${\bf
q}^{\prime}(r)$ is along the asymptotic momentum transfer ${\bf
q}$ which defines the ${\hat z}$-axis, and neglect the dependence
on ${\bf r}$ in the phases and in ${\bf q}^{\prime}(r)$, when
taking the divergence of ${\bf N}$. With this further
approximation, current conservation implies ${\omega}N_{0}+{\bf
q}^{\prime}(r){\cdot}{\bf N}=0$. Using these results, approximate
the cross section  for the inclusive reaction $(e,e')$ can be
written as
\begin{equation}
\frac{d^2\sigma}{d\Omega_e d\omega}= \sigma_{M} \{
\frac{q^4_\mu}{q^4}  S^{'}_{L}(q,w) + [ \tan^2 \frac{\theta_e}{2}
- \frac{q^2_\mu}{2q^2} ]  S^{'}_{T}(q,w) \}
\end{equation}
and the transform of the transition nuclear current elements which
appears in $S_{L}$ and $S_{T}$ are given by
\begin{eqnarray}
N^{appro. DW}_{0}&=&\int ({\frac {q'_{\mu}(r)} {q_{\mu}}})^{2}
({\frac {q} {q'(r)}})^{2}e^{i{\delta}_{f}([{\bf r}{\times} {\bf
p}_{i}^{\prime}(r)]^{2})}e^{i{\delta}_{f}([{\bf r}{\times} {\bf
p}_{f}^{\prime}(r)]^{2})}e^{i({\Delta}_{i}-{\Delta}_{f})}
e^{i{\bf q}^{\prime}(r){\cdot}{\bf r}}J_{0}({\bf r})d^{3}r
\label{apn0}\\
{\bf N}^{appro. DW}_{T}&=&\int e^{i{\delta}_{i}([{\bf
r}{\times} {\bf p}_{i}^{\prime}(r)]^{2})}e^{i{\delta}_{f}([{\bf
r}{\times} {\bf
p}_{f}^{\prime}(r)]^{2})}e^{i({\Delta}_{i}-{\Delta}_{f})}
e^{{\imath}{\bf q}^{\prime}(r){\cdot}{\bf r}}{\bf
J}_{T}({\bf r})d^{3}r \label{apnt}
\end{eqnarray}

Due to the angular dependence in the phase factors in
Eqs. (\ref{apn0}) and (\ref{apnt}), a multipole expansion of the
approximate potential is not practical.  Thus, $N^{appro. DW}_0$
and ${\bf N}^{appro. DW}_T$ have to be evaluated by carrying out a
3-dimensional numerical integration.  As we have shown for the
$(e,e'p)$ case \cite{kim3}, this numerical integration reproduces
the exact DWBA results very well.  However, since the inclusive
reaction $(e,e')$ requires a sum over all occupied neutron and
proton shells and a further integration over the directions of the
outgoing nucleon, numerical integration is very time consuming. In
order to have a more practical procedure we examine additional
approximations that will allow the integration over the angular
coordinates in Eqs. (\ref{apn0}) and (\ref{apnt}) to be done
analytically.

We created such an {\em ad-hoc} procedure in a previous paper
\cite{kim1}, but we were comparing our {\em ad-hoc} procedures to
the exact DWBA calculation which was largely dominated by the
transverse terms.  Hence, our {\em ad-hoc} procedures for the
longitudinal term were not very well determined.  In addition,
our full DWBA calculation was only set up for electrons, so we
could not check the {\em ad-hoc} approximation for positrons.  In
order to address this matter, we created a simple {\em toy} model
which assumes harmonic oscillator bound state protons and takes
the outgoing continuum proton wavefunction to be a plane wave.
Using this simple model to calculate the transition charge
distributions allows us to calculate the longitudinal
contribution to the cross section using the approximate DW
expression for $N_0$ of Eq. (\ref{apn0}) and to compare this
result to various {\em ad-hoc} proscriptions.  Based on this
investigation, coupled with our previous investigation of the
transverse contributions which dominate the cross section at
large electron scattering angles, we propose the following {\em
ad-hoc} expressions for the longitudinal and transverse structure
functions:

\begin{eqnarray}
N^{ad-hoc}_{0}&=& \int ({\frac {q'_{\mu}(r)} {q_{\mu}}})^{2}
({\frac {q} {q'(r)}})^{2}
e^{i<\delta(\kappa_i^2)+\delta(\kappa_f^2)>}e^{i{\bf
q}'(r){\cdot}{\bf r}}J_{0} ({\bf r})d^{3}r \label{ad-hoc0} \\
{\bf N}^{ad-hoc}_{T}&=&({\frac {p_{i}'(0)} {p_{i}}}) \int e^{i{\bf
q}'(r){\cdot}{\bf r}}{\bf J}_{T}({\bf r}) d^{3}r. \label{ad-hocT}
\end{eqnarray}
where $<\delta(\kappa_{i,f}^2)>$ denotes an average over the
angles of the vector ${\bf r}$.  That is, $<\kappa_{i,f}^2>=
<({\bf r} \times {\bf p}_{i,f})^2>=r^2 p^{2}_{i,f}(3-cos^2
\theta_{p_{i,f}})/4$.
  Note that under this averaging, the $\Delta$ term goes to zero.  This removes the
angular dependence in the phase factors, and thus permits a
multipole treatment of the matrix element as usual.

In the following figures for the longitudinal parts of the cross
sections based on our simple model we will compare our new
recommended longitudinal {\em ad-hoc} result given in Eq.
(\ref{ad-hoc0}) to the result calculated by the full
three-dimensional integration of Eq. (\ref{apn0}) and to our
previous {\em ad-hoc} results called LEMA$'$ which we give below
for convenience:
\begin{eqnarray}
N^{LEMA'}_{0}&=&(\frac{p_{i}^\prime(0)}{p_i} \int e^{i{\bf
q}''(r){\cdot}{\bf r}}J_{0} ({\bf r})d^{3}r \label{LEMA'}
\end{eqnarray}
where ${\bf q}''={\bf p}''_i(r) - {\bf p}''_f(r)$, $
p''(r)=p-\frac{\lambda}{r}\int_0^r V(r')dr'$ and the factor
$\lambda$, which depends on the energy transfer $\omega$, is given
by $\lambda=(\omega/\omega_o)^2$ with $\omega_o=\frac{q^2}{1.4
M}$.

Clearly $N^{ad-hoc}_{0}$ and ${\vec{\bf N}}^{ad-hoc}_{T}$
represent a modified Fourier transform of the nuclear transition
current. For comparison purposes, the approximation known as the
EMA replaces ${\bf q}'(r)$ with ${\bf q}'(0)$ wherever it appears
in Eqs.~(\ref{apn0}) and ~(\ref{apnt}) for $N_{0}$ and $N_{T}$ and
the phases are neglected as usual. We find that for light nuclei
the EMA is adequate, but it leads to large errors for nuclei as
heavy as $^{208}Pb$.

In Fig. 1  we compare the two approximate calculations with the
DW approximation for the longitudinal contribution to the cross
section for knocking protons out of various shells at a forward
angle in $^{208}Pb$ by electrons or positrons. Note that while we
use harmonic oscillator wavefunctions for all orbitals, we do use
the binding energies of the orbitals that correspond to the
values we find for our relativistic $\sigma-\omega$ model for
$^{208}Pb$.   While the {\em ad-hoc} result is not in perfect
agreement with the full DW result, it clearly is in better
agreement that the LEMA$'$ result and, for cases where the
electron incident and final energy exceed 300 MeV is in
reasonable agreement, particularly near the maxima.    Note that
the positron results are not very sensitive to which
approximation is used.

In Fig. 2 we show similar results at a backward angle.  We note that our
{\em ad-hoc} DWBA results for positrons tend to be in much better
agreement with the DW result than the electron case.  We again
find that our new {\em ad-hoc} approximation for the longitudinal
contribution is considerably better that our previous LEMA$'$
result.  We note that while the agreement between our {\em ad-hoc}
calculation and the  DW calculation for knocking out protons from
individual orbitals is not excellent, the discrepancies do not
seem have a systematic tendency to be either low or high and we
have reason to hope that when all the orbitals are added together
as in the case of $(e,e')$ reactions from nuclei that these
discrepancies will tend to average out.

\section{Comparison to experiment and conclusions}

Based of our investigation of this simple {\em toy} model, we
adopt our new {\em ad-hoc} model for the longitudinal structure
functions and return to our full nuclear model for investigating
Coulomb corrections for  $^{208}Pb(e,e')$ in the quasielastic
region where the lepton can be electrons or positrons.  Our first
step is to re-examine our full DWBA calculation \cite{jin2} and
modify the code for the case of positrons. We were successful in
doing this and can now compare the full DWBA calculation for
electrons and positrons based on a realistic relativisitic
nuclear model to our {\em ad-hoc} treatment of Coulomb
corrections which still permit a separation into longitudinal and
transverse terms.

In Fig. 3 we compare the full DWBA calculation to the {\em ad-hoc}
result and the electron plane wave result (PWBA) for electrons
with incident energy of $E_i=310$ MeV and scattering angle of
$\theta=143^o$. Note that this comparison is only a test of the
our {\em ad-hoc} transverse treatment which is unchanged from our
previous work since the longitudinal contribution at such a large
angle is at the few percent level. The agreement of the
"plane-wave-like" {\em ad-hoc} calculation with the full DWBA
result is quite good, even though the outgoing electron energy is
well below $300$ MeV.

In Fig. 4, we perform a similar comparison for the case of
positrons with incident energy of $E_i=485$ MeV and scattering
angle $\theta=60^o$.  Again, the agreement is quite good, and
unlike the backward angle scattering case, the longitudinal
response contributes about 40\% of the cross section.

We have examined a number of other cases, and the agreement shown
in Figs. 3 and 4 is characteristic at these energies.  As the
lepton energies increase, the {\em ad-hoc} approximation improves
since the Coulomb distortion effects become smaller. We did
notice in our investigations a general tendency that Coulomb
distortion effects for positrons tend to be smaller than Coulomb
distortion effects for electrons.  This corresponds to an
observation made many years ago when looking at inelastic lepton
scattering from nuclei \cite{lew}, where we noted that Coulomb
distortion for positrons tends to saturate.  As electrons pass
near the nucleus, the attactive Coulomb interaction pulls them
into regions of stronger potential which increases the Coulomb
distortion effects, while positrons are pushed away from the
region with a stronger potential.

With our capability of examining Coulomb distortion of both
positrons and leptons with the full DWBA calculation and with our
improved {\em ad-hoc} procedure we can compare our model
predictions to experiment.  In Fig. 5, we compare our model
calculations with Coulomb distortion included exactly and with
our {\em ad-hoc} method for quasielastic scattering of electrons
of energy $383$ MeV and positrons of energy $420$ MeV both at a
scattering angle of $\theta=60^o$ from $^{208}Pb$ to the
experimental data from Saclay \cite{gueye,morgenstern}. Note that
in this and the following figure, we are plotting  the total
structure function $S_{total} = {\frac {d^2\sigma} {d\omega
d\Omega_f}}/{\sigma_{M}(E_i)}$.

We first note that our {\em ad-hoc} and exact DWBA results are in
reasonable agreement although the lepton energy is somewhat low
for our approximate result, and further that the positron and
electron total structure functions have approximately the same
shape as a function of the energy transfer $\omega$.  However,
they do not have the same magnitude as do the data from Saclay.
The positron theory result is in reasonable agreement with the
experimental data, but the electron result is approximately
15\%-20\% larger than the data.

In Fig. 6 we make a similar comparison except that now the
scattering angle is $\theta=143^o$, and the electron incident
energy is $224$ MeV while the positron incident energy is $262$
MeV.  Again, when $S_{total}$ is plotted the positron and electron
shapes as a function of energy transfer $\omega$ are very
similar, but again, unlike the experimental data, the magnitudes
are quite different.  At this backward scattering angle case, our
electron result (DWBA) is in quite good agreement with the data.
At these much lower energies, clearly our {\em ad-hoc}
approximation is beginning to fail, particularly for the electron
case.

There is considerable interest in extracting the longitudinal
contributions from $(e,e')$ reactions from medium and heavy
nuclei in order to investigate the Coulomb sum rule.  Clearly,
Coulomb distortion effects have to be handled properly.  Our
results indicate that we could use a {\em Rosenbluth-like}
procedure in order to separate our ``longitudinal'' and
``transverse'' contributions to the cross section.  However,
these contributions depend on a modified (by Coulomb distortion)
Fourier transform of the transition charge and current
distributions.  It is necessary to use a nuclear model to extract
the longitudinal and/or tranverse structure functions from the
data.  It is not clear to us that a {\em Rosenbluth-like}
procedure is the best way to proceed, since our {\em ad-hoc}
procedure is not accurate in the wings of the cross section
distributions and in many cases, some of the Rosenbluth points
fall on either the low $\omega$ or high $\omega$ side of the
quasielastic peak.  It seems that a better procedure might be to
choose some semi-realistic nuclear model for the process in
question.  Use Eqs. (\ref {ad-hoc0}) and (\ref {ad-hocT}) to
calculate the structure functions and then fit the calculations
to the available data using a least squares procedure to
determine nomalization factors $N_L$ and $N_T$ in front of the
appropriate terms. The nuclear model should have the overall
correct spatial and kinematic dependence, but the longitudinal or
transverse strength will be determined by fitting these
normalization factors.  Once these factors are determined, one
can use the same nuclear model weighted with these factors to
calcuate the plane wave structure functions, thereby having
``measured'' the nuclear longitudinal and transverse response.

\section*{Acknowledgments}

This work was supported in part by the U.S. Department of Energy
under Grant No. FG02-87ER40370 and the Korean Ministry of
Education through BK21 Physics Research Division.  We thank
Professor Morgenstern for providing us the Saclay data files.

\begin{figure}[p]
\caption[fig1]{Longitudinal contributions to the differential cross
sections at a forward scattering angle for
${^{208}}Pb(e_{\pm},e'_{\pm})$ for different bound state
orbitals. The solid line is the approximate DW result, the dashed
line is our {\em ad-hoc} result and the dotted  line is our
previous LEMA$'$ approximation.} \label{fig1}
\end{figure}

\begin{figure}[p]
\caption[fig2]{Longitudinal contributions to the differential cross
sections for ${^{208}}Pb(e_{\pm},e'_{\pm})$ at a backward angle
for different bound state orbitals. The solid line is the
approximate DW result, the dashed line is our {\em ad-hoc} DWBA
result and the dotted  line is our previous LEMA$'$
approximation.} \label{fig2}
\end{figure}

\begin{figure}[p]
\caption[fig3]{The DWBA differential cross section for
${^{208}}Pb(e_{-},e'_{-})$ at $310$ MeV and scattering angle
$\theta = 143^o$ compared to our {\em ad-hoc} DWBA and to the
plane wave result. The bound state and continuum neutron and
proton orbitals are solutions to relativistic Hartree potential
based on the $\sigma-\omega$ model.} \label{fig3}
\end{figure}

\begin{figure}[p]
\caption[fig4]{The DWBA differential cross section for
${^{208}}Pb(e_{+},e'_{+})$ at $485$ MeV and scattering angle
$\theta = 60^o$ compared to our {\em ad-hoc} DWBA and to the plane
wave result. The bound state and continuum neutron and proton
orbitals are solutions to relativistic Hartree potential based on
the $\sigma-\omega$ model.}
\label{fig4}
\end{figure}

\begin{figure}[p]
\caption[fig5]{The total structure function $S_{total}$ generated by
dividing the differential cross section by $\sigma_M$ for
$^{208}Pb(e_{\pm},e'_{\pm})$ at a forward scattering angle of
$60^o$ with electrons of energy $383$ MeV and positrons with
energy $420$ MeV. The theoretical curves correspond to the full
DWBA calculation and to our {\em ad-hoc} DWBA calculation. The
data were taken at Saclay \cite{gueye,morgenstern}. The bound
state and continuum neutron and proton orbitals are solutions to
relativistic Hartree potential based on the $\sigma-\omega$
model.} \label{fig5}
\end{figure}

\begin{figure}[p]
\caption[fig6]{The total structure function $S_{total}$ generated by
dividing the differential cross section by $\sigma_M$ for
${^{208}}Pb(e_{\pm},e'_{\pm})$ at a backward scattering angle of
$143^o$ with electrons of energy $224$ MeV and positrons with
energy $262$ MeV. The theoretical curves correspond to the full
DWBA calculation and to our {\em ad-hoc} DWBA calculation. The
data were taken at Saclay \cite{gueye,morgenstern}. The bound
state and continuum neutron and proton orbitals are solutions to
relativistic Hartree potential based on the $\sigma-\omega$
model.} \label{fig6}
\end{figure}


\newpage

\begin{references}
\bibitem{jin1}Yanhe Jin, D. S. Onley, and L. E. Wright, Phys. Rev.
{\bf C45}, 1311(1992).
\bibitem{yates} C.F. Williamson, T.C. Yates, W. M. Schmitt, M. Osborn,
M. Deady, Peter, D. Zimmerman, C.C. Blatchley, Kamal K. Seth, M. Sarmiento,
B. Parker, Yahne Jin, L.E. Wright and D.S. Onley, Phys. Rev. C{\bf 56},
3152(1997).
\bibitem{jin2}Yanhe Jin, D. S. Onley, and L. E. Wright, Phys. Rev.
{\bf C45}, 1333(1992).
\bibitem{zhang}  Yanhe Jin, J.K. Zhang, D.S. Onley and L.E. Wright,
Phys. Rev. C{\bf 47}, 2024(1993).
\bibitem{jin3} Yanhe Jin, D.S. Onley and L.E. Wright, Phys. Rev. C{\bf 50},
168(1994).
\bibitem{horo}C. J. Horowitz and B. D. Serot, Nucl. Phys.
{\bf A368}, 503(1981).
\bibitem{hama}S. Hama, B. C. Clark, E. D. Cooper, H. S. Sherif,
and R. L. Mercer, Phys. Rev. C{\bf 41}, 2737(1990).
\bibitem{gent}V. Van der Sluys, J. Ryckebush, and M. Waroquier, Phys. Rev.
{\bf C54}, 1322(1996): Phys. Rev. {\bf C55}, 1982(1997).
\bibitem{kim1}K. S. Kim, L. E. Wright, Yanhe Jin, and D. W. Kosik,
Phys. Rev. {\bf C54}, 2515(1996).
\bibitem{kim2}K. S. Kim and L. E. Wright, Phys. Rev. {\bf C56},
302(1997).
\bibitem{kim3} K. S. Kim and L. E. Wright, Phys. Rev.{\bf C}60, 067604 (1999).

\bibitem{knoll}J. Knoll, Nucl. Phys. {\bf A201}, 289 (1973):
{\bf A223}, 462(1974).
\bibitem{lenz}F. Lenz and R. Rosenfelder, Nucl. Phys. {\bf A176},
513(1971); F. Lenz, thesis, Freiburg (1971).
\bibitem{giusti}C. Giusti, and F. D. Pacati, Nucl. Phys. {\bf A473},
717(1987).
\bibitem{trani}M. Trani, S. Turck-Chieze and A. Zghiche, Phys.
Rev. {\bf C38}, 2799 (1988).
\bibitem{lew} D.S. Onley, J.T. Reynolds and L.E. Wright,
Phys.\ Rev.\ {\bf 134}, B945 (1965).
\bibitem{gueye} P. Gu\`{e}ye, et al., Phys.\ Rev.\ {\bf C60}, 044308
(1999).
\bibitem{morgenstern} J. Morgenstern, Private Communication.

\end{references}
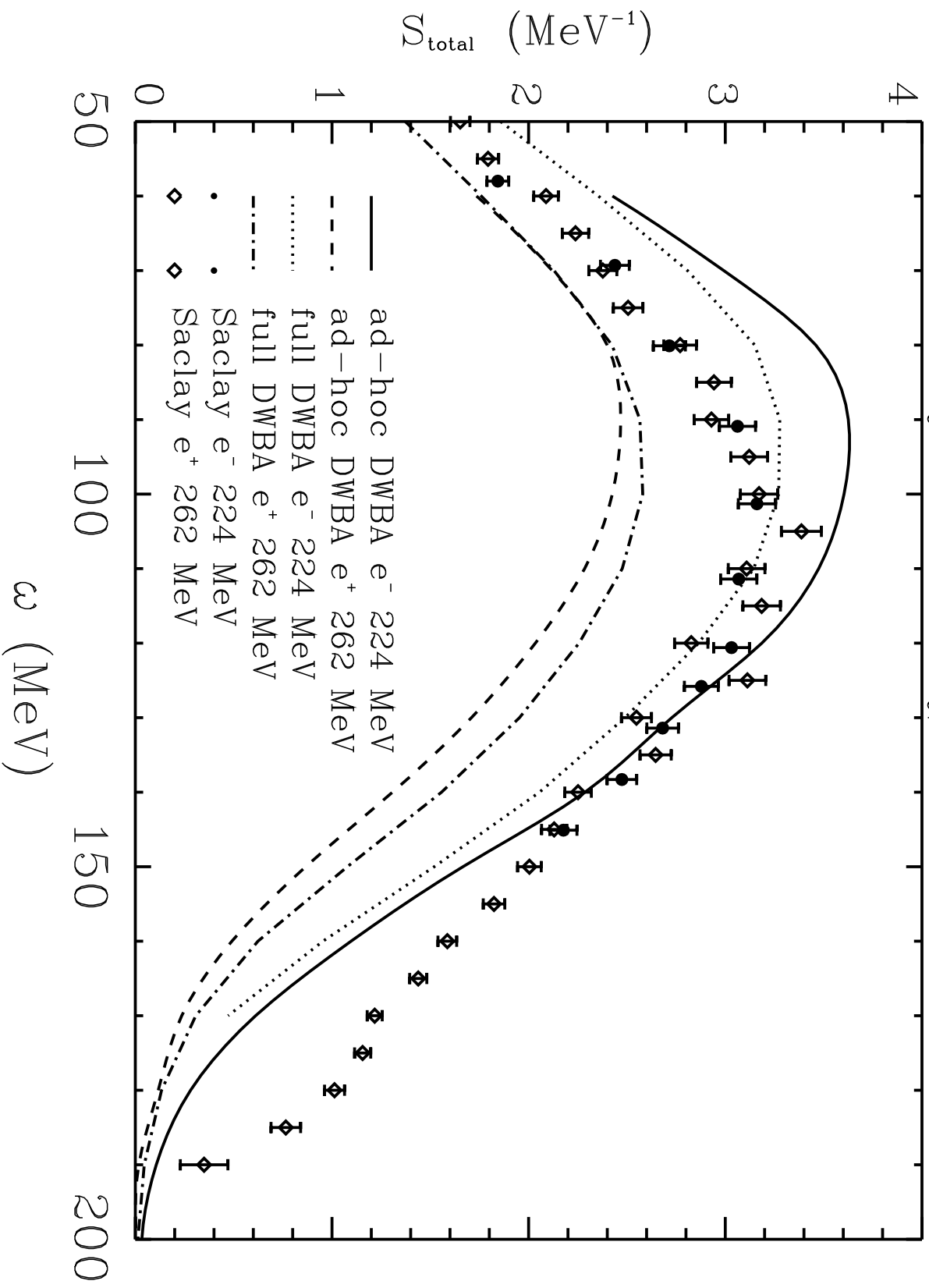
\end{document}